\documentclass[final]{aipproc}
\usepackage{graphicx,amsmath,amsfonts,amssymb,mathrsfs}

\layoutstyle{8x11double}

\newcommand{\eg}{{\it e.g.}}

\newcommand{\etc}{{\it etc.}}

\newcommand{\fig}{Figure}

\newcommand{\Ref}{Ref.}
\newcommand{\Refs}{Refs.}

\newcommand{\stheta}{\sin^22\theta_{13}}

\newcommand{\figu}[1]{\fig~\ref{fig:#1}}

\newcommand{\bi}{\begin{itemize}}
\newcommand{\ei}{\end{itemize}}

\begin{document}

\title{Physics and Performance Evaluation Group}

\keywords{Neutrino Oscillations, Neutrino Factory
\hfill
{\bf Preprint:} IPPP/07/95, DCPT/07/190 
}

\author{Andrea Donini}{
  address={Instituto F{\'i}sica Te{\'o}rica UAM/CSIC, Cantoblanco, E-28049 Madrid, Spain}
}

\author{Patrick Huber}{
  address={Physics Department, Theory Division, CERN, CH-1211 Geneva 23, Switzerland}
}

\author{Silvia Pascoli}{
  address={IPPP, Department of Physics, Durham University, Durham DH1 3LE, United Kingdom}
}

\author{\\Walter Winter}{
  address={Institut f{\"u}r theoretische Physik und Astrophysik, Universit{\"a}t W{\"u}rzburg, 
D-97074 W{\"u}rzburg, Germany}
}

\author{Osamu Yasuda}{
  address={Department of Physics, Tokyo Metropolitan University, Minami-Osawa, Hachioji, Tokyo 192-0397, Japan}
}

\begin{abstract}
We summarize the objectives and results of the ``international 
scoping study of a future neutrino factory and superbeam facility'' (ISS) physics working group.
Furthermore, we
discuss how the ISS study should develop into a neutrino factory design study (IDS-NF) from the point of view
of physics and performance evaluation.
\end{abstract}

\maketitle

\renewcommand{\thefootnote}{\fnsymbol{footnote}}

\footnotetext[1]{E-Mail: {\tt andrea.donini@uam.es}}
\footnotetext[2]{E-Mail: {\tt pahuber@vt.edu}}
\footnotetext[3]{E-Mail: {\tt silvia.pascoli@durham.ac.uk}}
\footnotetext[4]{Speaker, E-Mail: {\tt winter@physik.uni-wuerzburg.de}}
\footnotetext[5]{E-mail: {\tt yasuda@phys.metro-u.ac.jp}}


\section{Introduction}

It is the main objective of any future neutrino factory, beta beam, 
or superbeam neutrino oscillation facility to provide information on $\stheta$,
the neutrino mass hierarchy, and leptonic CP violation.
These and other observables,  such 
as deviations from maximal atmospheric neutrino mixing,
turn out to be indications in favor of theories of lepton masses and mixings (see, \eg, 
\Refs~\cite{Antusch:2004yx,Albright:2006cw}). In addition,
a detection of leptonic CP violation might be a good
motivation to suspect the origin of the baryon asymmetry
of the universe in the lepton sector. 
Any information from any such future neutrino oscillation facility can furthermore provide hints for new physics.

\section{ISS summary}

The physics case for a future accelerator-based neutrino oscillation 
facility was therefore made in
\Ref~\cite{ISS} within the framework of the ``International 
scoping study of a future neutrino factory and superbeam facility'' (ISS)
(for earlier studies, see \Refs~\cite{Apollonio:2002en,Albright:2004iw}).
The physics working group of this year-long study focused on 
``establishing the physics case for the various facilities and in finding the optimum parameters of the accelerator facility and detector systems from the physics point of view''.
It consisted of four major subgroups. The theory subgroup
identified the big questions which are essential for the physics case, such
as questions related to the origin of neutrino mass, the role of neutrinos
in the early universe, information from neutrinos about unification of 
forces and matter, \etc. The phenomenology subgroup discussed searches for clues of new physics 
in neutrino  facilities alone, and in combination with other
experiments. The experimental subgroup performed a detailed comparison of the performance of 
the various facilities. The muon physics subgroup reviewed muon physics that can be studied 
with the high intensity muon beam available 
at a neutrino factory.

The key objective of the ISS report~\cite{ISS} is to present the first
detailed comparison of the performance of various facilities.
The following represents the executive summary results with respect to the
neutrino factory. 
Using realistic specifications, the likely performance is estimated, 
optimum combinations of facilities,
baselines, and neutrino energies, are tried to be found, and some
staging scenarios are attempted to be identified.
Although the neutrino factory can achieve very large data samples with
small backgrounds, it operates at energies considerably higher than
the first oscillation peak $(E_{\mathrm{max}}/{\mathrm{GeV}}=L/564 ~{\rm km})$.
Because of this, at intermediate values of 
$\theta_{13}$ ($10^{-3} \lesssim \sin^2 2 \theta_{13} \lesssim 10^{-2}$)
the neutrino factory with only one golden-channel 
($\nu_e\to \nu_{\mu}$ and $\bar{\nu}_e\to \bar{\nu}_{\mu}$) detector (at, say, 4000~km) can not 
resolve all parameter degeneracies and the precision of the
measurement of a particular parameter is reduced by correlations 
among the parameters. 
These problems can be resolved in one of three ways:
\begin{enumerate}
  \item
    Placing a second detector at a different baseline (i.e. varying 
    the ratio $L/E$ );
  \item
    Adding a detector sensitive to the silver channel 
    ($\nu_e\to \nu_{\tau}$); or  
  \item
   Using an improved detector with lower neutrino-energy threshold and
   better energy resolution.
\end{enumerate}
Possible configurations for each alternative, alone and in
combination, were investigated to find an optimum performance of the
neutrino factory.
It was shown that a considerable reduction of parent muon-energy down
to $\sim 25$~GeV is feasible without a significant loss of
oscillation-physics output, provided a detector performance improved
with respect to the one assumed in earlier studies can be achieved.

\begin{figure}[t]
\includegraphics[width=0.9\columnwidth]{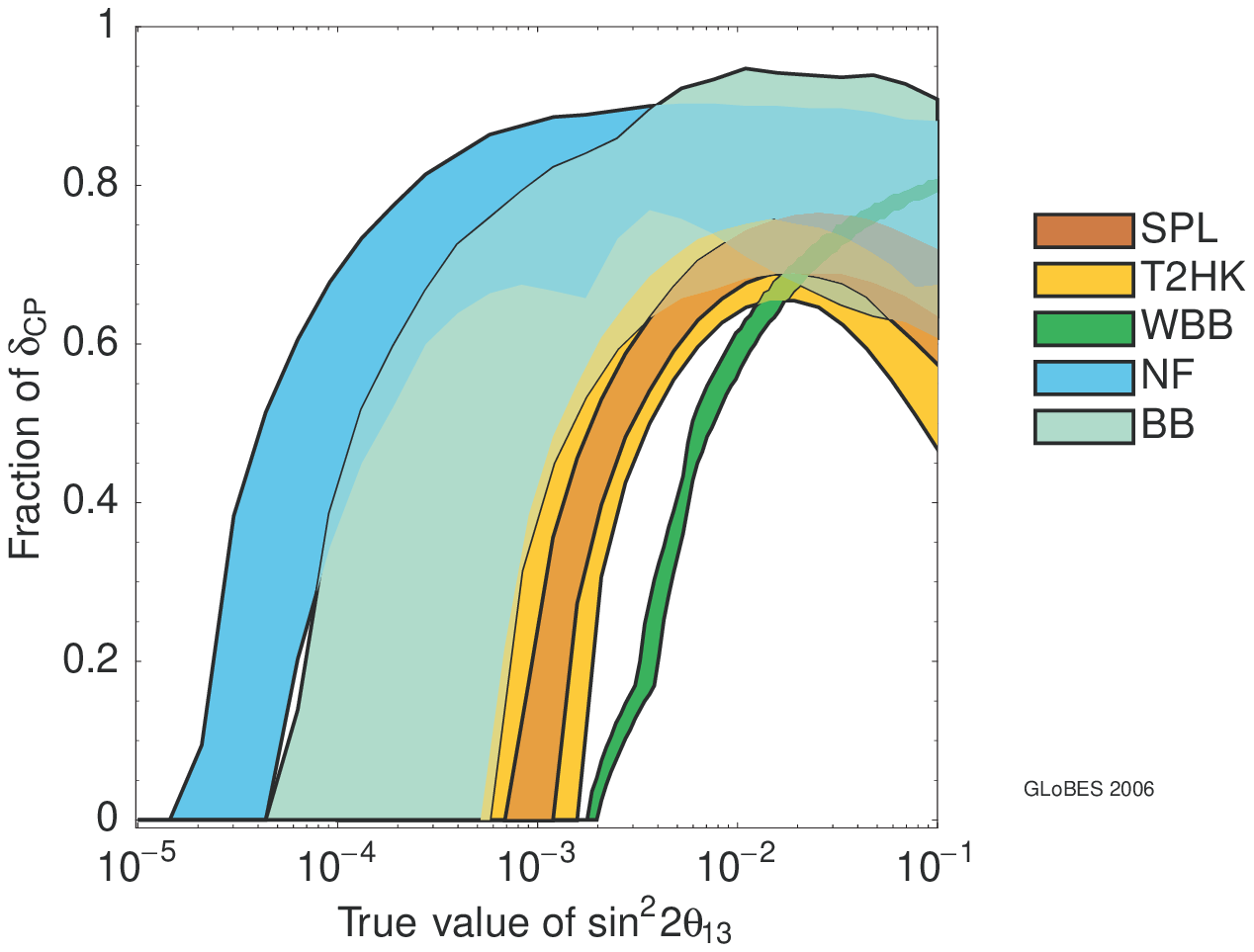}  \\
\includegraphics[width=0.9\columnwidth]{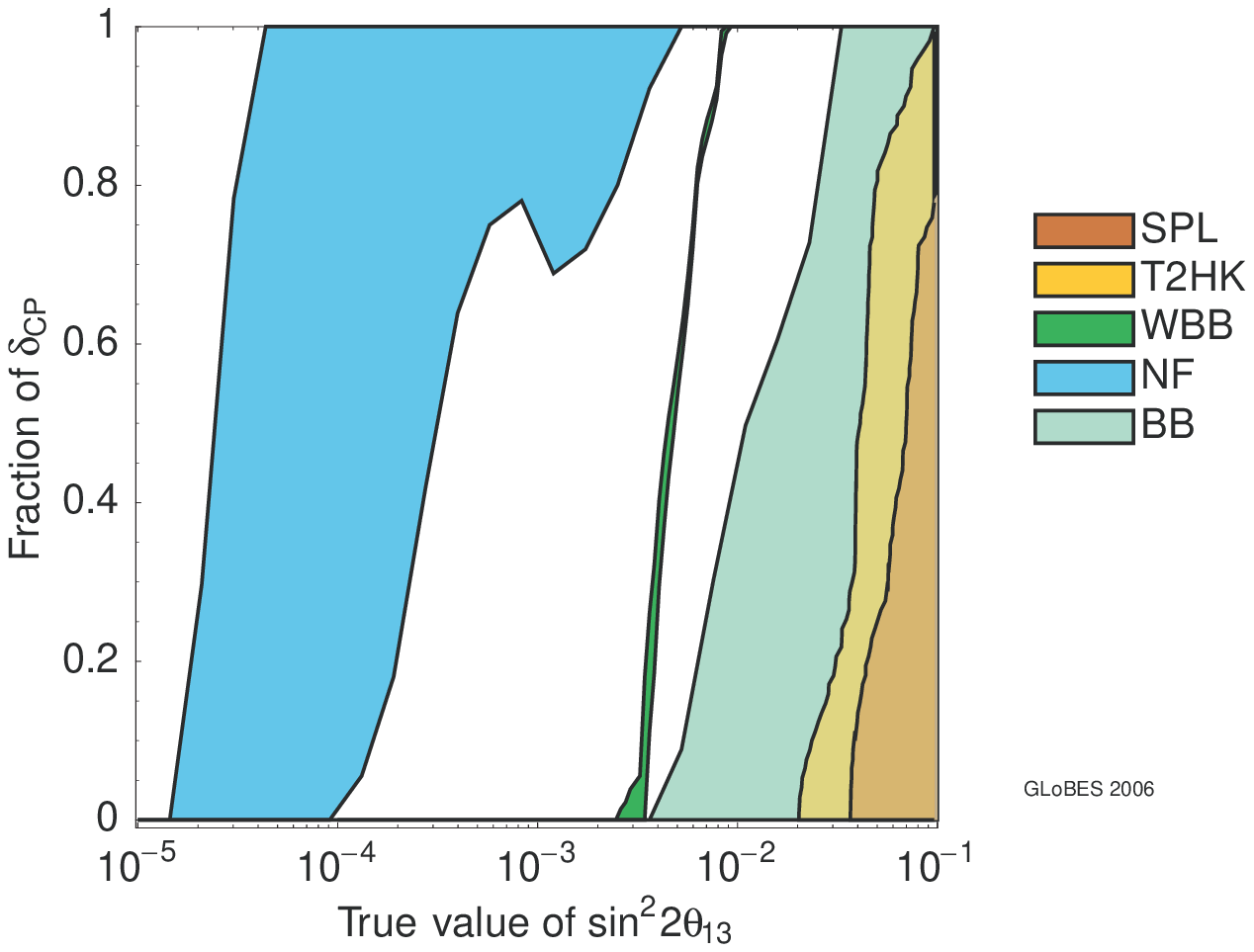} \\
\caption{\label{fig:cpv} The CP violation (left) and mass hierarchy (right) 
discovery reaches of 
various proposed facilities. The figure shows the fraction of all
possible values of $\delta$ for which CP violation or the mass hierarchy can be discovered
as a function of the simulated $\stheta$.
    The right-hand edges of the bands correspond to conservative
    setups while the left-hand edges correspond to optimized
    setups. Figure from \Ref~\cite{ISS} computed using the 
GLoBES software~\cite{Huber:2004ka,Huber:2007ji}.}
\end{figure}

As one of the results, we show in \figu{cpv} the CP violation and mass
hierarchy discovery reaches of 
various proposed facilities. The figure shows the fraction of all
possible values of $\delta$ for which CP violation (or the mass hierarchy)
can be discovered
as a function of the simulated $\stheta$. 
Three representative superbeam configurations were considered:
the SPL, a superbeam directed from CERN to the Modane laboratory;
T2HK, an upgrade of the J-PARC neutrino beam illuminating a detector
close to Kamioka, and the WBB, a wide-band, on-axis beam from BNL or
FNAL to a deep underground laboratory in the US. 
Each superbeam was assumed to illuminate a megaton-class water
Cherenkov detector. 
The beta beam options considered were the CERN baseline scheme, in
which helium and neon ions are stored with a relativistic $\gamma$ of
100, and an optimized beta beam, for which $\gamma=350$ (``BB'' in \figu{cpv}).
Two neutrino factory options were considered: a conservative option
with a single 50~kton detector sited at a baseline $\sim$2000~km - 4000~km 
from a 50~GeV Neutrino Factory; and the optimized Neutrino Factory (see the
full report) with two detectors, one at a baseline $\sim$2000~km - 4000~km  and
the second at the magic baseline~($\sim 7500$~km).
The result of the comparisons may be summarized as follows: 
for the options considered, the neutrino factory has the best
discovery reach for $\sin^2 2 \theta_{13}$ followed by the beta beam and
the superbeam, while the $\sin^2 2 \theta_{13}$ reach for resolving
the sign of the atmospheric mass difference is mainly controlled by
the length of the baseline.  
For large values of $\theta_{13}$ 
($\sin^2 2 \theta_{13} \gtrsim 10^{-2}$), the three classes of facility
have comparable sensitivity for the discovery of CP violation; the best
precision on individual parameters being achieved at the Neutrino
Factory using optimized detectors. 
The reduction of systematic uncertainties is the key issue at large
$\theta_{13}$; by reducing systematic uncertainties, the superbeam
may be favorably compared with the conservative neutrino factory.
For intermediate values of $\theta_{13}$ 
($10^{-3} \lesssim \sin^2 2 \theta_{13} \lesssim 10^{-2}$), the superbeams
are outperformed by the beta beam and the Neutrino Factory and the
best CP coverage is achieved by the beta beam. 
For small values of $\theta_{13}$
($\sin^2 2 \theta_{13} \lesssim 10^{-3}$), the neutrino factory
outperforms the other options.
Note, the comparisons are made using three performance indicators
only ($\sin^22\theta_{13}$, the sign of mass hierarchy and the CP
violating phase $\delta$). 
If other physics topics, such as the search for $e,\mu,\tau$ flavor
anomalies, were to be emphasized, the relative performance may be
different. 
From the physics point of view,
many of the different physics performance regions still overlap because
of yet unclear systematics, luminosities, \etc, which make a relative comparison
of different approaches very challenging. Therefore, from any future design
study, reliable predictions for these parameters as well as cost estimates
will be required. 

\section{IDS-NF perspectives}

As for a neutrino factory, an international design study (IDS-NF) will
continue the activities of the ISS. This follow-up study aims to present a design report, 
schedule, cost estimate, risk assessment for a neutrino factory. A part of this study
will be the ``physics and performance evaluation group'' (PPEG). At least in the first phase,
 the focus will be on the performance of the various neutrino facilities. At a later stage, the focus will be shifted towards the physics case for a neutrino factory, and requirements for muon physics
and non-standard physics will be included. The tasks of the PPEG
include the coordination of the physics performance study, the maintenance of a web site,
the interface with accelerator and detector working groups, possibly the organization
of workshops, and providing the necessary reports. 

After the ISS study, there are still many open questions to be addressed. For example,
the baseline design has to be fixed, in particular, in discussion with the detector 
working group. Furthermore, new ideas have to be evaluated. For example, for relatively large
$\stheta$, the possibility to have a low-energy neutrino factory has been drawing
some attention~\cite{Geer:2007kn,Bross:2007ts}, possibly in combination 
with a superbeam~\cite{Huber:2007uj}. Furthermore, the neutrino factory physics
potential has to be compared to the one of other experiment classes, such as beta beams.
The  requirements for non-standard physics measurements, such as the silver channel~\cite{Donini:2002rm},
have to be defined. Finally, the requirements for muon physics and how these interact with the oscillation program ought to be investigated.

The main purpose of the PPEG is to evaluate the physics performance of a given experimental setup (neutrino factory or other, existing or planned) in a transparent, consistent and documented fashion.
Hereby {\em transparent} means clearly stating the
definition of the performance indicator,
the approximations used, 
the chosen input parameters, the luminosity, 
the confidence level, and
the treatment of systematics. In addition,
{\em consistent} means that the  assumptions should be on equal footing, and {\em documented} means that the relevant information will be archived and will be accessible, possibly on a web site.
A major ingredient for this task is the definition of parameters.
For the neutrino factory setups, all the parameters will be defined in strict collaboration with the detector and accelerator working groups of the IDS. This requires a defined and efficient communication between the various working groups. There might be different types of setups:
\begin{enumerate}
\item
Baseline setup: stable, agreed design
\item
Conservative modification of 1.: for example, a different baseline
\item
Speculative ideas
\end{enumerate}
As for the beta beam setups, we will possibly follow the same procedure as for neutrino factory, working in collaboration with the beta beam groups. For the superbeam setups, the use of existing literature and input of well-established and recognized experts will be mandatory.

As far as the PPEG key ingredients are concerned, the main purpose will be performance evaluation
following the method used in the ISS report. This means the use of GLoBES software~\cite{Huber:2004ka,Huber:2007ji}.
Performance plots should be updated twice a year and presented at NuFact.
In Europe, there will be a close connection to the FP07 proposal, which is a 
 funding proposal for a high-energy neutrino oscillation facility design study
(in particular, between the IDS-NF PPEG and Euro$\nu$ design study Work Package~6: ``Physics reach and comparison'').
The PPEG has already set up a web site~\cite{IDSweb}.
It will serve to illustrate the mandate, structure and activities of the PPEG.
It will host the results of the performance evaluations. It will provide links
to the documentation relative to the performance evaluation. And it will 
inform about the activities with announcements and with links to the workshops, 
the reports and any other relevant activity.

\section{Summary}

In summary, the mandate of the PPEG is the continuation of the ISS activities, focusing 
on physics performance evaluation in a transparent, 
consistent, and documented. It is an international, inclusive effort, where everyone is invited 
to contribute.

\vspace*{0.3cm}

{\bf Acknowledgments} Some of us acknowledge the support of CARE,
contract number RII3-CT-2003-506395. We would also like to thank the organizers of NuFact 07 for the nice meeting.

\renewcommand{\refname}{}

\vspace*{-1.3cm}


\end{document}